# Estimation of WECC System Inertia Using Observed Frequency Transients

D. P. Chassin[2], Z. Huang[2], M. K. Donnelly[2], C. Hassler[1], E. Ramirez[1], C. Ray[1]

## ABSTRACT

Computer models being developed to understand the interaction between demand-response technology, power system deregulation and market transformation depend in part on understanding the relationship between system frequency and load-control. Frequency, load, and plant outage events data collected over the last several years have permitted analysis to determine the Western Electricity Coordination Council (WECC) system's inertia during each event. This data was used to evaluate the relationship of system inertia to total load, which is used to model system response to load curtailment programs in next generation power system simulations.

**Index Terms -** System Inertia Property, Grid-Friendly Appliances, Frequency Excursion, WECC

## I. INTRODUCTION

Pacific Northwest National Laboratory (PNNL) has developed the Power Distribution System Simulator (PDSS) as a part of a broader program to study the interaction of power markets, generation resources, transmission dynamics, distribution controls and end-use load-shedding technologies [1]. PDSS includes so-called grid-friendly appliance control strategies, which are triggered by frequency deviations as observed at the point of power delivery to the end-use appliance.

Power system frequency drops when supply from generators falls below the demand. When the change in frequency is large, the reduction in supply can trigger protection systems that may result in system separation, loss of load and customer outages. Therefore, studies of system response to generation outages generally consider the response of system frequency.

Inoue *et al.* [2] conducted research in this subject to estimate the inertia constant of a power system and evaluate the total on-line spinning-reserve requirements. A polynomial approximation with respect to time was fit to the wave form of the transients in estimating the inertia constant. This effort was made to estimate and evaluate the dynamic behavior of the system frequency in loss of generation or load. This would lead to a system aggregate model that can predict the frequency response to changes in supply-load mismatch. This procedure was done for 10 events in the 60 Hz system of Japan. The difference between the Japanese and WECC systems resulted in large differences in the values for the inertia constant *M*. However, they were unable to establish a statistically significant relationship between *M* and the load because of the small number of events that were analyzed.

Generator outages across the western electric grid occur regularly. Frequency drops below 59.950 Hz have occurred almost daily since May 2002 when a frequency transducer was installed at PNNL and data collection of WECC frequency began at 10 samples per second.

Based in part on the PNNL data, system models for PDSS are being developed to represent grid behavior with various load-control technology deployment scenarios. Lumped models of how frequency responsive demand affects total inertia on the system is beyond the scope of the tools currently being developed. However, it is necessary to incorporate an aggregate system response model into PDSS to study the sub-minute behavior of frequency-responsive loads in power systems. The letter summarizes the results of an effort to derive such a model for the WECC.

## II. METHODOLOGY

Inoue [2] describes a method of inertia estimation based on the study of under-frequencies following generation outages. Significant deviations in frequency occur when there is a large difference between supply and load. The general equation of inertia can be simplified by considering damping effects to be small during early onset of the event. Without a significant loss of information in the results, we use the relation

$$-\Delta P = M \cdot \frac{df}{dt} \qquad (1)$$

where $df/dt$ is the frequency change in Hz/s, $\Delta P$ is the power change (pu in system load base), $M$ is the inertial constant in pu.seconds [3]. The inertial property is in large part determined by how many and which generators are running at the time of the outage. During the first seconds after an outage, the frequency is determined almost exclusively by the inertial response, as illustrated for the June 14th, 2004 event in Fig. 1.

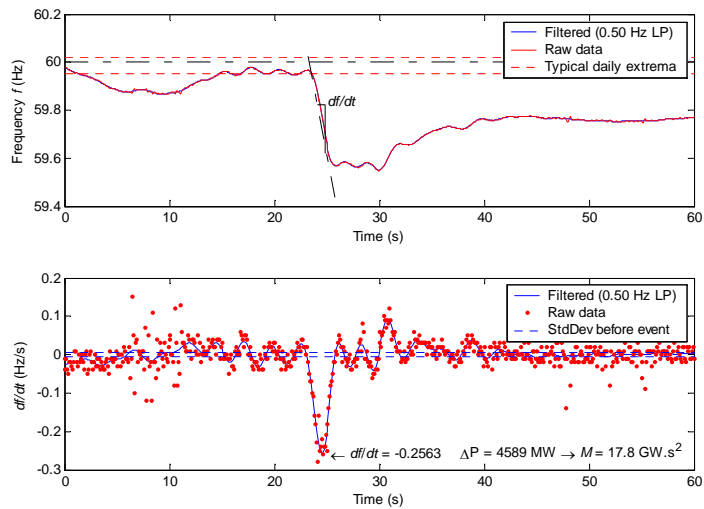

**Fig. 1:** Inertial analysis of June 14, 2004 event on WECC.

The frequency monitor deployed at PNNL collects WECC frequency at 10 samples per second with a 0.001 Hz resolution, as observed in Richland, Washington. The data was collected and archived from May 2002 through June 2004, with approximately 85% coverage over that time. The majority of the missing data is from the summer of 2002.

Information on 388 major plant outages during the study period was obtained from WECC. In addition, hourly system load data for all 11 WECC regions for the study period was obtained from PowerLytix, Inc.

The maximum frequency rate of change was estimated by removing noise in the original frequency data with a 0.5 Hz filter and computing the first derivative. The filtered $df/dt$ was then associated with power lost at each event. The result was compared with the WECC system load in an effort to determine the relationship between the inertial property and the power-load mismatch.

### III. RESULTS

The value of the inertial constant $M$ was successfully computed for 167 events, as shown in Fig. 2. The 221 remaining events have insufficient data to compute $M$ correctly at this time. The WECC reports for these events did not include the outage MW or the search algorithm was not able to obtain a frequency profile for the reported event time due to time skewing errors or excessive ambient noise in relation to the size of frequency deviation. Also any event that did not result in a frequency below 59.950 Hz was excluded. Finally, 6 light-load events with $M > 30$ seconds were also excluded.

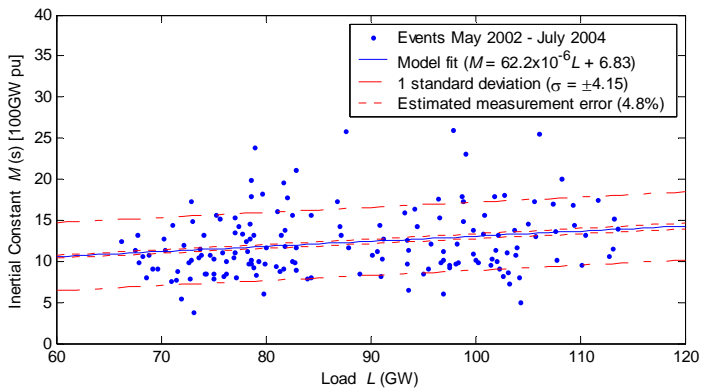

**Fig. 2: Inertia estimates for observed WECC outages (log scale).**

When the inertial constant was compared to system load, a linear fit was obtained. From this fit, we determined $M = 62.2 \times 10^{-6} L + 6.83$ seconds with one standard deviation at $\pm 4.15$ seconds.

### IV. DISCUSSION

There are a number of possible sources of error that should be recognized for future study using this method. Every effort has been made to eliminate human error in computing the frequency deviation and the effect of time-stamp skewing in the data. However, the accuracy of outage and load reports must also be considered.

The variation in inertia is greatly disproportionate to the overall range of inertia as load varies, suggesting that factors other than system load, such as seasonal variation of generator dispatch scheduling may contribute to variations in $M$ as well. However, we cannot exclude the consequence of inaccurate load and outage data for smaller plant outages.

The sources of inaccuracies must be considered before a reduced-order model of inertia can be implemented in PDSS. Three contributions to the errors in calculating $M$ are taken into consideration using the following equation:

$$\Delta M = \frac{\partial M}{\partial g} \Delta g + \frac{\partial M}{\partial f} \Delta f + \frac{\partial M}{\partial l} \Delta l \qquad (2)$$

where $g$ is the generation and $l$ is the load, both in MW, such that $P = g - l$. The accuracy of the data collection system was calculated for the June 14th event and estimated to contribute 0.43% error. Plant outages are reported in MW, but rounded to only 1 or 2 significant digits, which suggests that they are the scheduled output of the generation, and not the actual power output at the time of the outage. System load clearly varies during the hour from that forecasted, scheduled, or reported. Indeed, the observed frequency fluctuations in the data collected are caused principally by such "load-following" discrepancies. A typical 100 MW error in either load or generation would contribute 2.18% errors in the inertial constant.

Taken together, we estimate that the known sources of error contribute to about 4.8% of the observed error in $M$. Additional sources of error must be attributed to other unmodeled phenomena such as the non-uniform dispatch of generation assets with varying inertia, e.g., the annual variation of hydro generation, diurnal variations in generation dispatch to "peaker" plants.

### V. CONCLUSION

The data collected from WECC suggests that inertia is correlated to the overall system load but that significant errors remain in the reduced order model proposed for PDSS. Attempts to discern any time correlation were unsuccessful. A large number of events exist for the lower values of $M$, but there is a great deal of variation in the inertial constant. The rarity of large events significantly limits the accuracy of the estimated inertia for wide range of model conditions.

More precise methods of calculating the slope of the frequency deviation, as well as improved data filtering, could assure more accurate results of $M$. Not excluding the effects of damping is one possible solution for better accuracy that can be considered. Both values of $M$ and $D$ would need to be estimated using more sophisticated methods. Notwithstanding the difficulty, addressing these considerations would yield a more accurate model.

### VI. REFERENCES

[1] R. T. Guttromson, D. P. Chassin, and S. E. Widergren, "Residential energy resource models for distribution feeder simulation," IEEE Power Engineering Society General Meeting, vol. 1, 13-17 July 2003, pp. 108–113.
[2] T. Inoue, H. Taniguchi, Y. Ikeguchi, and K. Yoshida, "Estimation of power system inertia constant and capacity of spinning-reserve support generators using measured frequency transients," IEEE Transactions on Power Systems, vol. 12 no. 1 , Feb. 1997, pp. 136–143.
[3] P. Kundur, "Power System Stability and Control," *The EPRI Power System Engineering Series,"* McGraw-Hill, 1994. Section 11.1.